\documentclass[prd,aps,groupedaddress,showpacs,nofootinbib,amsmath]{revtex4}
\usepackage{epsfig}
\usepackage{color}
\usepackage{amssymb}
\usepackage{amsfonts}
\usepackage{graphicx}
\usepackage{keyval,graphicx}
\usepackage{textcomp,wasysym}
\usepackage{amsbsy}
\usepackage{subfigure}
\newcommand{\ud}{\mathrm{d}}
\makeatletter
  \newcommand\figcaption{\def\@captype{figure}\caption}
  \newcommand\tabcaption{\def\@captype{table}\caption}
  
  \newcommand{\Rmnum}[1]{\expandafter\@slowromancap\romannumeral #1@}
\makeatother

\begin{document}
\title{The mixing of $D_{s1}(2460)$ and $D_{s1}(2536)$}
\author{Xiao-Gang Wu$^1$}\email{wuxiaogang@ihep.ac.cn}
\author{Qiang Zhao$^{1,2}$}\email{zhaoq@ihep.ac.cn}

\affiliation{1) Institute of High Energy Physics, Chinese Academy of
Sciences, Beijing 100049, P.R. China }

\affiliation{2) Theoretical Physics Center for Science Facilities,
CAS, Beijing 100049, P.R. China }

\date{\today}

\begin{abstract}

The mixing mechanism of axial-vectors $D_{s1}(2460)$ and
$D_{s1}(2536)$ is studied via intermediate hadron loops, e.g. $D^*
K$, to which both states have strong couplings. By constructing the
two-state mixing propagator matrix that respects the unitarity
constraint and calculating the vertex coupling form factors in a
chiral quark model, we can extract the masses, widths and mixing
angles of the physical states. Two poles can be identified in the
propagator matrix. One is at $\sqrt{s}=2454.5 \ \textrm{MeV}$
corresponding to $D_{s1}(2460)$ and the other at
$\sqrt{s}=(2544.9-1.0i) \ \textrm{MeV}$ corresponding to
$D_{s1}(2536)$. For $D_{s1}(2460)$, a large mixing angle
$\theta=47.5^\circ$ between ${}^3P_1$ and ${}^1P_1$ is obtained. It
is driven by the real part of the mixing matrix element and
corresponds to $\theta'=12.3^\circ$ between the $j=1/2$ and $j=3/2$
state mixing in the heavy quark limit.  For $D_{s1}(2536)$, a mixing
angle $\theta=39.7^\circ$ which corresponds to $\theta'=4.4^\circ$
in the heavy quark limit is found. An additional phase angle
$\phi=-6.9^\circ \sim 6.9^\circ$ is needed at the pole mass of
$D_{s1}(2536)$ since the mixing matrix elements are complex numbers.
Both the real and imaginary part are found important for the large
mixing angle. We show that the new experimental data from BaBar
provide a strong constraint on the mixing angle at the mass of
$D_{s1}(2536)$, from which two values can be extracted, i.e.
$\theta_1=32.1^\circ$ or $\theta_2=38.4^\circ$. Our study agrees
well with the latter one. Detailed analysis of the mass shift
procedure due to the coupled channel effects is also presented.

\end{abstract}

\date{\today}



\maketitle

\section{Introduction}

In the past few years one of the most important experimental
progresses in the study of the charmed meson spectrum is the
establishment of the lowest $P$-wave charmed-strange mesons, i.e.
$D_{s0}(2317)$, $D_{s1}(2460)$, $D_{s1}(2536)$, and $D_{s2}(2573)$
as now listed in Particle Data Group (PDG) 2010
Edition~\cite{Nakamura:2010zzi}. Since the first observation by
BaBar Collaboration~\cite{Aubert:2003fg}, the spin-0 state
$D_{s0}(2317)$ and spin-1 $D_{s1}(2460)$ (later confirmed by
Belle~\cite{Krokovny:2003zq} and CLEO~\cite{Besson:2003cp}) have
initiated tremendous interests in its property and internal
structure. These two states have masses lower than the potential
model predictions, and their widths are rather narrow. It is somehow
agreed that their low masses are caused by the open $D K$ and $D^*K$
thresholds, respectively, and as a consequence, their narrow decay
widths are due to the dominant isospin-violating decays, i.e.
$D_{s0}(2317)\to D_s\pi$ and $D_{s1}(2460)\to D_s^*\pi$ (see the
review of Refs.~\cite{Swanson:2006st,Godfrey:2005ww}  and references
therein).

The heavy-light $Q\bar{q}$ system is an ideal platform for testing
the internal constituent quark degrees of freedom. In the heavy
quark limit the heavy quark spin is conserved and decoupled from the
light quark degrees of freedom, which are characterized by the total
angular momentum ${\bf j}_q\equiv {\bf s}_q+{\bf L}$, where ${\bf
s}_q$ is the light quark spin and ${\bf L}$ is its orbital angular
momentum. With $j_q=1/2$ and $j_q=3/2$, one can arrange those four
$P$-wave states into two classes, i.e. $J^P=0^+, \ 1^+$ and
$J^P=1^+, \ 2^+$, respectively, where $J$ is the meson spin as a sum
of the heavy quark spin ${\bf S}_Q$ and ${\bf j}_q$. For the axial
vector states in the charmed and charmed-strange meson spectrum,
since they are not charge conjugation eigenstates, state mixings
between the $^3P_1$ and $^1P_1$ configurations are allowed. In the
case of charmed and charmed-strange heavy-light system when the
heavy quark symmetry is broken at order of $1/m_c$, it would be
interesting to study the mechanism that causes deviations from the
ideal mixing scenario, i.e. breakdown of the heavy quark symmetry.
This forms our motivation in this work. As mentioned earlier,
$D_{s1}(2460)$ and $D_{s1}(2536)$  lie near the threshold of $D^* K$
and both couple to $D^* K$ strongly via a relative $S$ wave. It
gives rise to coupled channel effects in the mass shifts of
potential quark model calculations in comparison with the observed
values~\cite{Simonov:2004ar,Badalian:2007yr,Coito:2011qn}, and
produces state mixings between the $^3P_1$ and $^1P_1$
configurations. Similar mechanism has been studied in the
$a_0(980)$-$f_0(980)$ mixing in Ref.~\cite{Wu:2007jh}. Determination
of the mixing angle should be useful for understanding the property
and internal structure of these two axial vector states.

We mention that various solutions have been proposed in the
literature to explain the observed results for $D_{s1}(2460)$ and
$D_{s1}(2536)$. For instance,  $D^*K$ molecule or tetra-quark
configuration have been investigated in
Refs.~\cite{Close:2005se,Barnes:2003dj,Faessler:2007us}. In
Ref.~\cite{Guo:2006rp}, $D_{s1}(2460)$ is explained as a dynamically
generated state. The mixing angle has also been calculated in the
quark model~\cite{Godfrey:1985xj,Yamada:2005nu} but with large
uncertainties from the quark spin-orbital interactions. In this
work, we investigate the two-state mixing propagator matrix which
respects the unitarity constraint in a chiral quark model. We will
show that the coupled channel effects via intermediate hadron loops
can provide a simultaneous determination of the masses, widths and
mixing angles of these two axial vector states. We also mention that
the coupled channel effects on the ${}^3P_1$ and ${}^1P_1$ mixing
was recently studied in Ref.~\cite{Zhou:2011sp}, where the the
couplings were extracted in the ${}^3P_0$ model and  a subtracted
dispersion relation was applied to evaluate the hadron loops. In our
approach we use the chiral quark model to extract the couplings and
vertex form factor. We then extend the quark model form factor to a
covariant form which can be applied on a general ground to much
broader cases.

The paper is organized as follows. In Sec.~\ref{sec:cce}, we give
the basic formulas of two-state mixings via coupled channel
propagators. In Sec.~\ref{sec:vertices}, the relevant coupling form
factors are determined by the chiral quark model. In
Sec.~\ref{sec:propagator} the propagator matrix is calculated in
detail. Section~\ref{sec:mass} is devoted to show our numerical
results for the mass and mixing parameters. The experimental
constraints for the mixing angle are presented in
Sec.~\ref{sec:exp}. A summary is given in the last Section. In
Appendix~\ref{app:1} the detailed definition and calculation of a
special function used in the evaluation of the loop integrals with
exponential form factors are provided.

\section{Mixing through coupled channel effect}\label{sec:cce}

\begin{figure}[htbp]
  \includegraphics[scale=0.8]{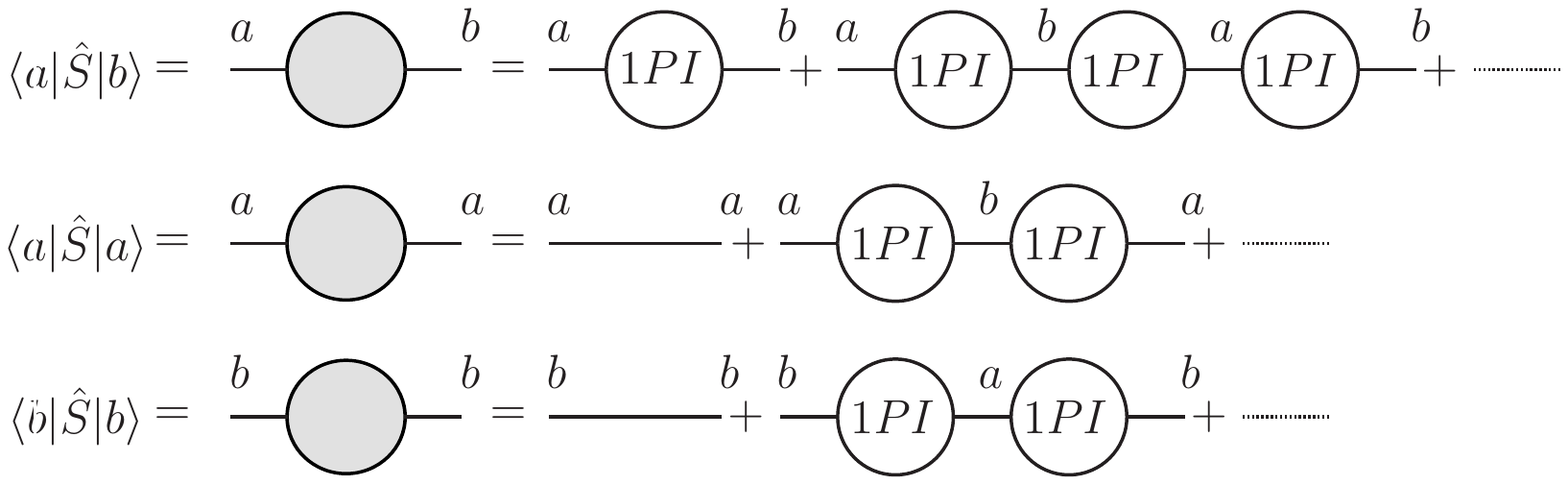}
  \vspace{-17.5cm}
  \caption{Transition through intermediate states}\label{chain}
\end{figure}

We use $|a\rangle$ and $|b\rangle$ to present two pure states in the
quark model. If they can couple to common final states, there will
be a transition between them via single particle irreducible (1PI)
diagrams as shown in Fig.~\ref{chain}.

The propagator matrix of $|a \rangle$ and $|b\rangle$ can be expressed as
\begin{equation}\label{eqn1}
    G_{ab}=\left(\begin{array}{c}
              \langle a |\\
              \langle b |
            \end{array}
            \right)
    \hat{S}
     \left( |a\rangle , |b\rangle
     \right) \ .
\end{equation}
The physical states $|A\rangle$ and $|B\rangle$ should be a mixture of $|a \rangle$ and $|b\rangle$,
\begin{equation}\label{eqn2}
    \left(\begin{array}{c}
            |A\rangle \\
            |B\rangle
          \end{array}
    \right)
    =
    \left(
    \begin{array}{cc}
      \cos \theta              & - \sin \theta e^{i \phi}\\
      \sin \theta e^{-i \phi}   & \cos \theta
    \end{array}
    \right)
    \left(\begin{array}{c}
            |a\rangle \\
            |b\rangle
          \end{array}
    \right)
    =
    R(\theta,\phi)
    \left(\begin{array}{c}
            |a\rangle \\
            |b\rangle
          \end{array}
    \right)
\end{equation}
where $R(\theta,\phi)$ is the mixing matrix,  $\theta$ is the mixing
angle, and $\phi$ is a possible relative phase between $|a \rangle$
and $|b\rangle$. Then the propagator matrix of $|A\rangle$ and
$|B\rangle$ is
\begin{equation}\label{eqn3}
G_{AB}=R G_{ab} R^{\dagger} \ .
\end{equation}
The physical propagator matrix $G_{AB}$ should be a diagonal matrix.
Thus, we can determine the mixing parameters $\{\theta,\phi\}$ by
diagonalizing the propagator matrix $G_{ab}$.

In the present case,  we set $|a\rangle=| {}^3P_1 \rangle$,
$|b\rangle=| {}^1P_1 \rangle$, $|A\rangle=|D_{s1}(2460)\rangle$ and
$|B\rangle=|D_{s1}(2536) \rangle$ as in Ref.~\cite{Badalian:2007yr}.
The mixing scheme is
\begin{eqnarray}\label{eqn2:1}
  |D_{s1}(2460)\rangle &=& \cos\theta | {}^3P_1 \rangle - \sin\theta e^{i \phi} | {}^1P_1 \rangle  \nonumber \\
  |D_{s1}(2536)\rangle &=& \sin\theta e^{-i\phi} | {}^3P_1 \rangle + \cos\theta | {}^1P_1
  \rangle \ ,
\end{eqnarray}
where states $| {}^3P_1 \rangle$ and $| {}^1P_1 \rangle$ can be
rotated to the eigenstates in the heavy quark limit:
\begin{equation}\label{eqn2:2}
    \left(\begin{array}{c}
    | {}^3P_1 \rangle \\ |{}^1P_1 \rangle
    \end{array}\right)=
    \left(\begin{array}{cc}
    \sqrt{\frac{2}{3}} & \sqrt{\frac{1}{3}} \\
    -\sqrt{\frac{1}{3}} & \sqrt{\frac{2}{3}}
    \end{array}\right)
    \left(\begin{array}{c}
    | j=\frac{1}{2} \rangle  \\ | j=\frac{3}{2} \rangle
    \end{array}\right) \ .
\end{equation}
The mixing angle $\theta$ defined in Eq.~(\ref{eqn2:1}) can be
related to $\theta'$ defined in $j=1/2$ and $j=3/2$ bases:
\begin{equation}\label{eqn2:3}
    \theta=\theta'+35.26^\circ \ .
\end{equation}

Considering parity conservation, the important intermediate states
that can couple to $D_{s1}(2460)$ and $D_{s1}(2536)$ are $D^* K$,
$D_{s}^{*} \eta$ and $D K^*$, of which the thresholds are listed in
Table~\ref{tab1}.

\begin{table}[htbp]
  \centering
  \caption{The thresholds of intermediate states for $D_{s1}(2460)$ and $D_{s1}(2536)$.}\label{tab1}
  \begin{tabular}{c|c|c|c|c|c}
  \hline\hline
  Intermediate states & $D^{*0} K^+$ & $D^{*+} K^0$  & $D_{s}^{*} \eta$ & $D^0 K^{*+}$ & $D^+ K^{*0}$\\
  \hline
  Threshold (GeV)       & 2.501         & 2.508         & 2.660            & 2.756        & 2.761   \\
  \hline\hline
\end{tabular}
\end{table}

\begin{figure}[htbp]
  \includegraphics[scale=0.7]{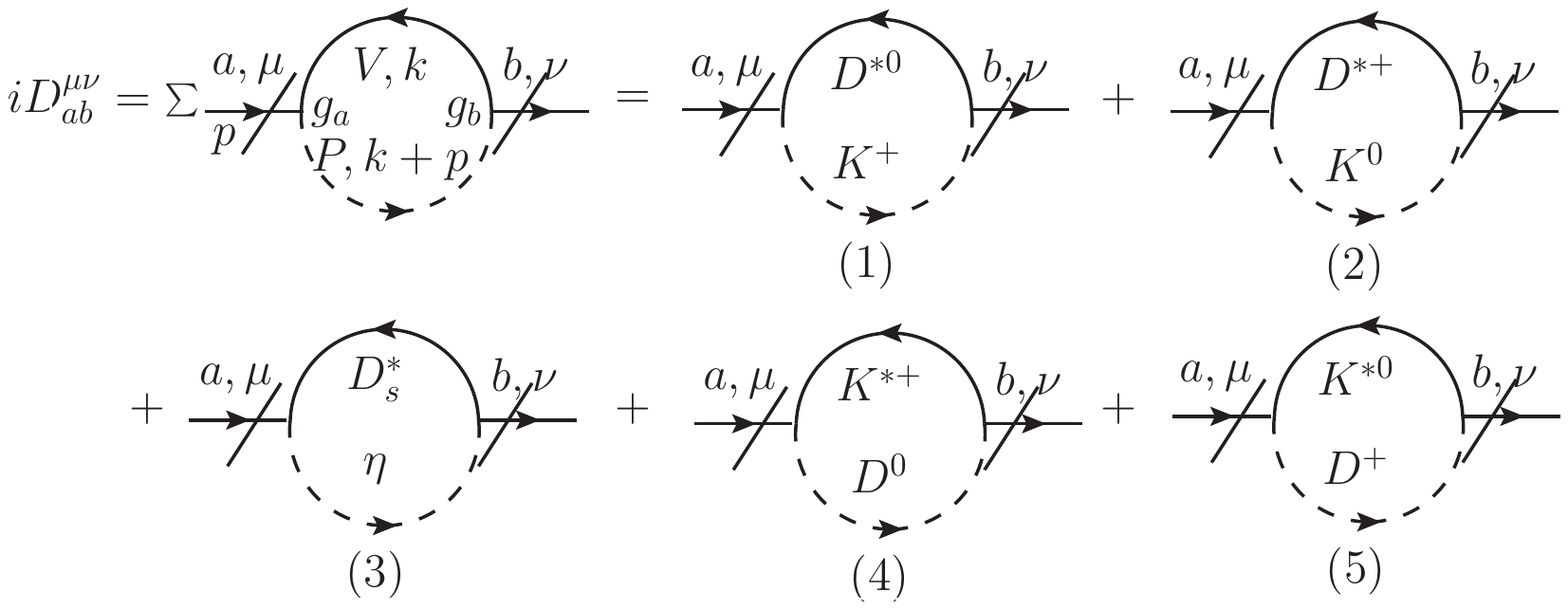}\\
  \vspace{-14cm}
  \caption{Mixing term for $D_{s1}(2460)$ and $D_{s1}(2536)$.}\label{mixterm}
\end{figure}

If all the particles involved are scalars or pseudoscalars,
Fig.~\ref{chain} will only represent sums of infinite geometric series
and the resulting propagator matrix $G$ becomes \cite{Wu:2007jh}
\begin{equation}\label{eqn4}
    G_{ab}=\frac{1}{D_a D_b-D_{ab}^2} \left(\begin{array}{cc}
                                            D_b & D_{ab} \\
                                            D_{ba} & D_a
                                          \end{array}
    \right) \ ,
\end{equation}
where $D_a$ and $D_b$ are the denominators of the single propagators
of $|a\rangle$ and $|b\rangle$, respectively, and the mixing term
$D_{ab}$ is the sum of all 1PI diagrams, which satisfies
$D_{ba}$=$D_{ab}$. But from Table \ref{tab1}, we find that the
particles involved in the present case can be scalars, vectors or
axial-vectors. There are five diagrams for the mixing of
$D_{s1}(2460)$ and $D_{s1}(2536)$ as shown in Fig. \ref{mixterm}.

The mixing term can be generally divided into transverse and
longitudinal terms:
\begin{equation}\label{eqn5}
    D_{ab}^{\mu \nu}\equiv  \Pi_{ab} P^{\mu \nu} + B_{ab} Q^{\mu \nu} \ ,
\end{equation}
where $P^{\mu \nu}\equiv g^{\mu \nu}-p^{\mu}p^{\nu}/p^2$ and $Q^{\mu
\nu}\equiv p^{\mu}p^{\nu}/p^2$ are the transverse and longitudinal
projector, respectively, and satisfy
\begin{equation}\label{eqn5a}
    P^{\mu \nu}P_\nu^{\lambda}=P^{\mu \lambda}, \
    Q^{\mu \nu}Q_\nu^{\lambda}=Q^{\mu \lambda}, \
    P^{\mu \nu}Q_\nu^{\lambda}=0 \ .
\end{equation}

Next we concentrate on the evaluation of the propagator matrix
$G^{\mu \nu}$ for axial vector states. The numerator of the vector
propagator is $g^{\mu \nu}-p^{\mu}p^{\nu}/m^2$ and can be generally
expressed as $P^{\mu \nu}+ \Delta Q^{\mu \nu}$ where
$\Delta=1-p^2/m^2$. With the properties of Eq.(\ref{eqn5a}), the
geometric sums, e.g. $\langle a | \hat{S} | b\rangle$ in
Fig.~\ref{chain}, can be taken for the transverse and longitudinal
terms independently. After include the self-energy functions
$\Pi_a^{\mu \nu}$ and $\Pi_b^{\mu \nu}$, the complete propagator
matrix for the $1{}^3 P_1$ and $1{}^1 P_1$ states becomes
\begin{equation}\label{eqn6}
    G_{ab}^{\mu \nu}=iP^{\mu \nu}
    \frac{\bar{G}_{ab}(s)}{\det{\bar{G}_{ab}}(s)}+i Q^{\mu \nu} \frac{G^L_{ab}}{\det{G^L_{ab}}} \ ,
\end{equation}
with
\begin{equation}\label{eqn7}
    \bar{G}_{ab}(s)\equiv M_{ab}^2-\delta_{ab} s=
    \left(\begin{array}{cc}
       m_b^2+\Pi_b(s)-s     & -\Pi_{ab}(s) \\
       -\Pi_{ab}(s)         & m_a^2+\Pi_a(s)-s
    \end{array}\right) \ ,
\end{equation}
and
\begin{equation}\label{eqn7a}
    G^L_{ab}(s)=\left(\begin{array}{cc}
    \frac{m_b^2-s}{\Delta_b}+B_b(s)  & -B_{ab}(s) \\
    -B_{ab}(s)                       & \frac{m_a^2-s}{\Delta_a}+B_a(s)
    \end{array}\right) \ ,
\end{equation}
where $M_{ab}^2$ is the mass matrix. After diagonalization, the mass
matrix becomes
\begin{equation}\label{eqn8}
    M_{AB}^2=R M_{ab}^2 R^{\dagger}=
    \left(\begin{array}{cc}
            m_B^2 & 0 \\
            0     & m_A^2
          \end{array}
    \right) \ .
\end{equation}
Note that the longitudinal term $G^L_{ab}/\det{G^L_{ab}}$ is nonvanishing,
but the poles are only related to the transverse term $\bar{G}_{ab}$.

By searching for the poles in the propagator matrix $G^{\mu
\nu}(s)$, which is equivalent to set $\det{\bar{G}(s)}=0$, we can
obtain the masses and widths of the physical states. In general,
there are two solutions $s_A$ and $s_B$ for the two state system. We
can also extract the mixing angle $\theta_{A,B}$ and the relative
phase angle $\phi_{A,B}$. These mixing parameters are different for
these two states, since they are extracted at the physical masses of
these two states, respectively. If $G$ is a normal matrix, which
means $G G^{\dagger}=G^{\dagger}G$, then it can be diagonalized
through a unitary transformation $R$. The resulting mixing angle
$\theta$ and relative phase $\phi$ can thus be uniquely determined.
Otherwise, we can only get a quasi-diagonalized matrix through the
unitary transformation $R$. The reason is because that orthogonality
cannot be satisfied between these two physical states.

\section{Coupling form factors in the chiral quark model}\label{sec:vertices}

\begin{figure}[htbp]
  \includegraphics[scale=0.7]{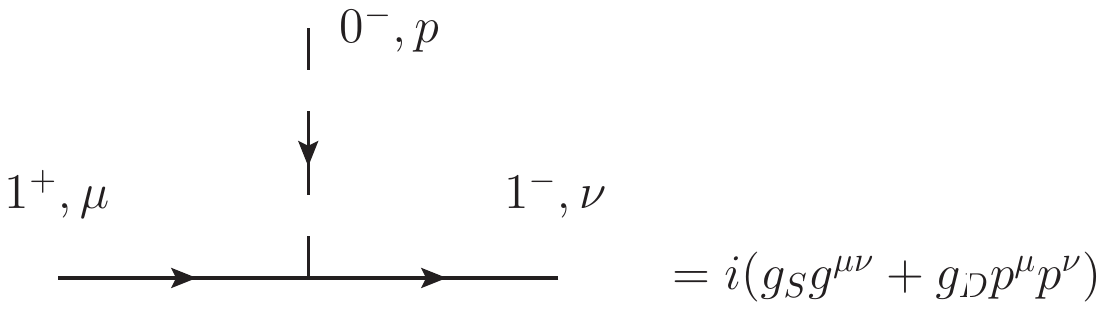}\\
  \vspace{-17cm}
  \caption{The AVP vertex via the $S$ and $D$ wave couplings.}\label{vertex}
\end{figure}

At hadronic level all the vertices in Fig.~\ref{mixterm} involve the
Axial-Vector-Pseudoscalar (AVP) type of coupling. In general,
the AVP coupling vertex contains two coupling constants $g_S$ and
$g_D$ representing the $S$ and $D$ waves as shown in
Fig.~\ref{vertex}.

Since the decay momentum is small near the threshold, we expect that
contributions from the $D$-wave coupling would be small. As a
reasonable approximation, we omit $g_D$ and keep $g_S$ to the order
$O(v^0)$. In the multipole approach, the helicity amplitude for $1^+
\rightarrow 1^- + 0^-$ takes the form \cite{Zou:2002yy}
\begin{equation}\label{eqn9}
 A_{\nu}=\langle S_f,\nu;0,0| \hat{S} | S_i,\nu \rangle = \sum_L \langle L, 0; S_f,\mu | S_i, \nu \rangle Y_{L0}(\hat{q})
 G_L \ ,
\end{equation}
where $G_L$ is the coupling constant for the $L$ wave and $\hat{q}$
is the momentum direction of the final state particle in the center
of mass frame of the initial state. In the present case,
Eq.~(\ref{eqn9}) becomes
\begin{equation}\label{eqn10}
    \left[
    \begin{array}{c}
      A_0 \\
      A_1
    \end{array}
    \right]
    =
    \left[\begin{array}{cc}
            \frac{1}{2 \sqrt{\pi}} & -\frac{1}{\sqrt{2 \pi}} \\
            \frac{1}{2 \sqrt{\pi}} & \frac{1}{2 \sqrt{2 \pi}}
          \end{array}
    \right]
    \left[\begin{array}{c}
            G_S \\
            G_D
          \end{array}
    \right] \ .
\end{equation}
In order to obtain $g_S$ to the order $O(v^0)$, we set
\begin{equation}\label{eqn11}
    g_S=A_0(\vec{q} \to 0) =A_1(\vec{q} \to 0) \ .
\end{equation}
\begin{figure}[htbp]
  \includegraphics[scale=0.7]{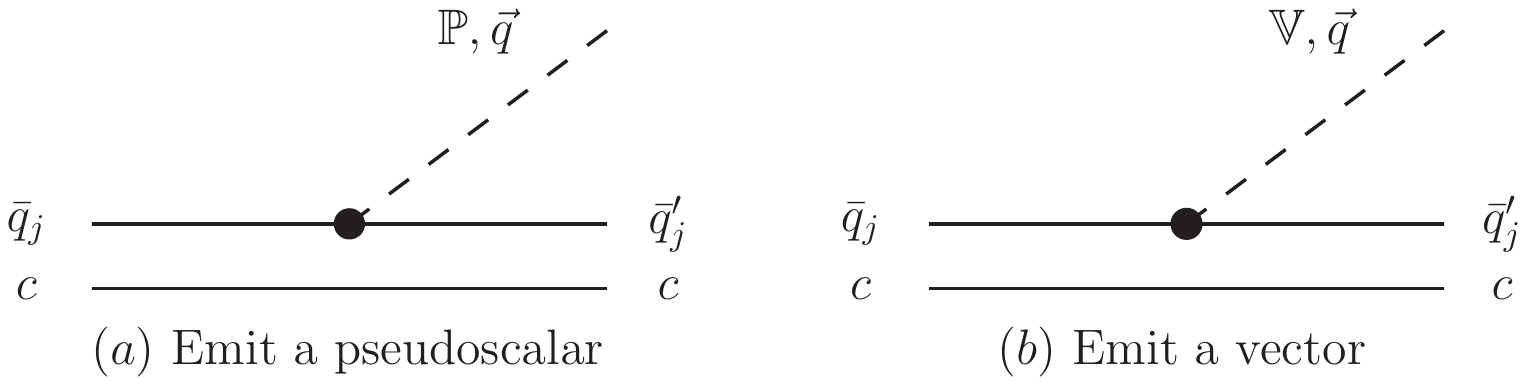}\\
  \vspace{-16cm}
  \caption{Pseudoscalar (a) and vector meson (b) emission via an active light quark $j$ in an effective chiral quark model.}\label{pic4}
\end{figure}

\subsection{Coupling to $D^* K$ and $D_{s}^{*} \eta$}

One notices that at all the coupling vertices the interacting quarks
involve only light quark, i.e. $u$, $d$ and $s$. By treating the
light mesons, pseudoscalar and vector mesons, as induced fields by a
chiral Lagrangian for the mesons coupling to constituent
quarks~\cite{Manohar:1983md}, the light and heavy quark degrees of
freedom can be separated out in terms of nonrelativistic expansions
near the decay threshold. This approach has been successfully
applied to light meson productions in photo-nucleon and
meson-nucleon
scatterings~\cite{Li:1997gd,Zhao:1998rt,Zhao:1998fn,Zhao:2002id,Zhao:2000iz,Zhong:2007fx,Zhong:2011ti}
and strong decays of heavy-light
mesons~\cite{Zhong:2008kd,Zhong:2009sk} recently.

In the chiral quark model, we treat the pseudoscalar mesons $K$ and
$\eta$ as the effective chiral fields as shown in
Fig.~\ref{pic4}(a). For emitting a pseudoscalar from an active quark
line, the quark-meson coupling and corresponding non-relativistic
form are respectively as follows~\cite{Zhong:2008kd}:
\begin{eqnarray}
  \label{eqn12.1}
  H_m &=& \sum_j \frac{1}{f_m} \hat{I}_j \bar{\psi}_j \gamma^{j}_{\mu} \gamma^{j}_{5} \psi_{j} \partial^{\mu} \phi_m \ ,\\
  \label{eqn12.2}
  H^{nr}_{m} &=& \sum_j \frac{1}{f_m}
                \left\{
                     G \boldsymbol{\sigma}_j \cdot \textbf{q} +  h \boldsymbol{\sigma}_j \cdot \textbf{p}^{i'}_j
                \right\}
                \hat{I}_j \exp(-i \textbf{q} \cdot \textbf{r}_j) \ ,
\end{eqnarray}
with
\begin{equation}\label{eqn13}
    G\equiv -\left(1+\frac{\omega}{E_f+M_f}\right),
    \quad
    h\equiv \frac{\omega}{2 \mu_q} \ ,
\end{equation}
where $f_m$ is the decay constant of the pseudoscalar meson,
$\hat{I}_j$ the isospin operator, $\omega$ the energy of the
pseudoscalar, $M_f$ and $E_f$ the mass and energy of the final state
heavy meson, $\mu_{q}$  a reduced mass given by $1/\mu_{q}\equiv
1/m_j+1/m_j'$, $\textbf{p}^{i'}_j$ and $\textbf{r}_j$ the internal
momentum and coordinate for the light ($j$th) quark of the final
state heavy meson.

Following the procedure in \cite{Zhong:2008kd},  we derive the
helicity amplitude $A^{q}_{\nu}\equiv \langle S_f, \nu| \hat{H}_m
|S_i, \nu \rangle$ in the quark level. For $1{}^3P_1 \to 1{}^3S_1 +
\mathbb{P}$, the explicit expressions are
\begin{equation}\label{eqn14}
    A^q_0=i g_1 h \alpha \exp(-\frac{q_1^2}{4 \alpha^2}),\quad
    A^q_1=i \frac{g_1}{4 \alpha} \left[2 G q q_1 + h (4 \alpha^2-q_1^2)\right] \exp(-\frac{q_1^2}{4
    \alpha^2}) \ ,
\end{equation}
and for $1{}^1P_1 \to 1{}^3S_1 + \mathbb{P}$, we have
\begin{equation}\label{eqn15}
    A^q_0=-i \frac{g_1}{2\sqrt{2} \alpha}\left[2G q q_1 + h (2\alpha^2-q_1^2) \right] \exp(-\frac{q_1^2}{4 \alpha^2}),\quad
    A^q_1=- \frac{i}{\sqrt{2}}g_1 h \alpha \exp(-\frac{q_1^2}{4
    \alpha^2}) \ ,
\end{equation}
where $g_1=\langle \mathbb{M}_f | \hat{I}_1| \mathbb{M}_i \rangle$
is the isospin factor,  $\alpha$ the harmonic oscillator strength
$\alpha\equiv \beta \left( {2 m_2}/(m_1+m_2) \right)^{{1}/{4}}$ as
in Ref.~\cite{Zhong:2008kd}, and $q_1\equiv q m_2/(m_1+m_2)$. In the
$c\bar{s}$ system, the 1st quark is $\bar{s}$  and the 2nd is $c$
quark, and the flavor symmetry between the heavy and light quark is
apparently broken.

By taking equivalence between the quark and hadron level helicity
amplitudes, we can extract the coupling form factor as follows:
\begin{equation}\label{eqn16}
A_{\nu}=\sqrt{(E_i+M_i)(E_f+M_f)} A^{q}_{\nu} \ .
\end{equation}
Then from Eqs.(\ref{eqn11}), (\ref{eqn14}) and (\ref{eqn15}), we
finally obtain:
\begin{eqnarray}
  \label{eqn17.1}
  \textrm{for}\quad 1{}^3P_1, & & g_S=-\frac{\delta}{f_m} \sqrt{2M_i(E_f+M_f)} \cdot  g_1 h \alpha \exp(-\frac{q_1^2}{4 \alpha^2}) \ ,\\
  \label{eqn17.2}
  \textrm{for}\quad  1{}^1P_1, & & g_S= \frac{\delta}{f_m} \sqrt{2M_i(E_f+M_f)} \cdot \frac{1}{\sqrt{2}} g_1 h \alpha \exp(-\frac{q_1^2}{4
  \alpha^2}) \ ,
\end{eqnarray}
where $\delta$  is a global parameter accounts for the strength of
the quark-meson couplings as introduced in \cite{Zhong:2008kd}.

\subsection{Coupling to $D K^*$}

In this coupling, the vector meson $K^*$ is treated as an effective
chiral field, for which the effective quark-vector-meson coupling
Lagrangian and the corresponding non-relativistic coupling
form~\cite{Zhao:1998fn,Riska:2000gd} are
\begin{eqnarray}
  \label{eqn18.1}
  \hat{H}_v &=& \sum_j a \bar{\psi}_j \gamma_{\mu}^j \phi_v^{\mu} \psi_j  \ ,\\
  \label{eqn18.2}
  \hat{H}_v^T &=& \sum_j \left\{ -\frac{\textbf{p}_j^{i'} \cdot \boldsymbol{\epsilon}^*}{2 \mu_q} + i \boldsymbol{\sigma}_j \cdot \textbf{q} \times \boldsymbol{\epsilon}^* \left( \frac{1}{2 m_j}+\frac{1}{E_f+M_f}-\frac{m_j'}{2 M' m_j} \right)
                       \right\} a \hat{I}_j \exp(-i \textbf{q} \cdot \textbf{r}_j) \ ,\\
  \label{eqn18.3}
  \hat{H}_v^L &=& \sum_j \left\{ \left[ \frac{q}{\mu}(1-\frac{\omega}{2 m_j})+ \frac{q \omega}{2 M' \mu} + \frac{q \omega m_j'}{2 M' \mu m_j} \right]-\frac{\omega}{2 \mu \mu_q} \textbf{p}_j^{i'} \cdot \hat{\textbf{q}}
  \right\} a \hat{I}_j \exp(-i \textbf{q} \cdot \textbf{r}_j) \ ,
\end{eqnarray}
where $\mu$, $\omega$ and \textbf{$\epsilon$} are the mass, energy
and  polarization vector of the emitted vector meson,  $M'$ the sum
of the constituent quark mass of the final meson, $a$ the overall
quark-vector-meson coupling, and other symbols have the same meaning
as those in Eqs.(\ref{eqn12.1})-(\ref{eqn13}). Using the above
operators, we can extract the helicity amplitudes $A_{\nu}^q$, i.e.
for $1{}^3P_1 \to 1{}^1S_0 + \mathbb{V}$,
\begin{equation}\label{eqn19}
A_1^q=-\frac{i}{2 \sqrt{2}} a g_1 B \frac{q_1}{\alpha}
\exp(-\frac{q_1^2}{4 \alpha^2}) ,\quad A_0^q= 0 \ ,
\end{equation}
and for $1{}^1P_1 \to 1{}^1S_0 + \mathbb{V}$,
\begin{equation}\label{eqn20}
A_1^q= i a g_1 A \alpha \exp(-\frac{q_1^2}{4 \alpha^2}) ,\quad
A_0^q= \frac{i}{\sqrt{2}}  a g_1 \left[ - C \frac{q_1}{\alpha} + D
\alpha \left(\frac{q_1^2}{2 \alpha^2} -1\right) \right]
\end{equation}
with
\begin{eqnarray}
  \label{eqn21.1}
  A &\equiv & -\frac{1}{2\sqrt{2} \mu_q} \ ,\\
  \label{eqn21.2}
  B &\equiv & -\sqrt{2} q \left( \frac{1}{2 m_j} + \frac{1}{E_f + M_f} - \frac{m_j'}{2M' m_j} \right) \ ,\\
  \label{eqn21.3}
  C &\equiv & -\left[ \frac{q}{\mu}(1-\frac{\omega}{2 m_j})+ \frac{q \omega}{2 M' \mu} + \frac{q \omega m_j'}{2 M' \mu m_j} \right] \ ,\\
  \label{eqn21.4}
  D &\equiv & \frac{\omega}{2 \mu \mu_q} \ .
\end{eqnarray}
Substituting Eqs.~(\ref{eqn19}) and (\ref{eqn20}) to
Eqs.~(\ref{eqn11}) and (\ref{eqn16}),
we obtain
\begin{eqnarray}
  \label{eqn22.1}
  \textrm{for} \quad 1{}^3P_1, && g_S=0 \ ,\\
  \label{eqn22.2}
  \textrm{for} \quad 1{}^1P_1, && g_S=- \sqrt{(E_i+M_i)(E_f+m_f)}  \cdot \frac{1}{\sqrt{2}} a g_1 \frac{\alpha}{\mu_q} \exp(-\frac{q_1^2}{4
  \alpha^2}) \ .
\end{eqnarray}
The coupling $g_S=0$ in Eq.~(\ref{eqn22.1}) is because $A_1^q$ in
Eqs.~(\ref{eqn19}) is proportional to $q_1$. Thus, the effective
coupling vanishes below the open decay threshold. As a consequence,
the contributions from Fig.~\ref{mixterm}(4) and (5) should vanish
to the order $O(v^0)$. Therefore, we only need to consider the
contributions from Fig.~\ref{mixterm}(1-3) in the following
calculations.

\subsection{Numerical results for couplings}

In the numerical calculation,  we set $|\vec{q}|=0$ when the initial
state lies below the threshold for $\mathbb{V P}$ final state
\cite{Riska:2000gd}. We adopt
$\eta=\frac{1}{\sqrt{3}}(u\bar{u}+d\bar{d}-s\bar{s})$ in the
$\eta-\eta'$ mixing scheme which corresponds to
$\theta_P=-\arcsin(1/3)=-19.47^\circ$ for the flavor octet and
singlet mixing. Since the contribution from the $D_s^*\eta$ loop is
small, the uncertainties with $\theta_P$ have only negligible
effects on the mixing matrix. We obtain isospin factors $g_1$ for
different intermediate states as listed in Table~\ref{tab2}.
\begin{table}[htbp]
  \centering
  \caption{Isospin factors $g_1$ extracted in the quark model.}\label{tab2}
  \begin{tabular}{c|c|c|c}
    \hline\hline
    $1{}^3P_1 / 1{}^1P_1$ & $D^{*0} K^+$ & $D^{*+} K^0$ & $D_s^{*} \eta$ \\
    \hline
    $g_1$                 & 1            & 1          & $-\frac{1}{\sqrt{3}}$ \\
    \hline\hline
  \end{tabular}
\end{table}

The following values are adopted for other
parameters~\cite{Zhong:2008kd}: $\delta=0.557$, $\beta=0.4 \
\textrm{GeV}$, $f_K=f_{\eta}=160 \ \textrm{MeV}$, and the
constituent quark masses $m_u=m_d=350 \ \textrm{MeV}$, $m_s=550 \
\textrm{MeV}$, $m_c=1700 \ \textrm{MeV}$. We note that our numerical
results are not sensitive to $m_c=1500\sim 1700$ MeV, while the
light quark masses $m_u=m_d=330\sim 350$ MeV and $m_s=500\sim 550$
MeV will lead to about $(1\sim 5)\%$ uncertainties with the final
results.

\begin{figure}[htbp]
  \includegraphics[scale=1]{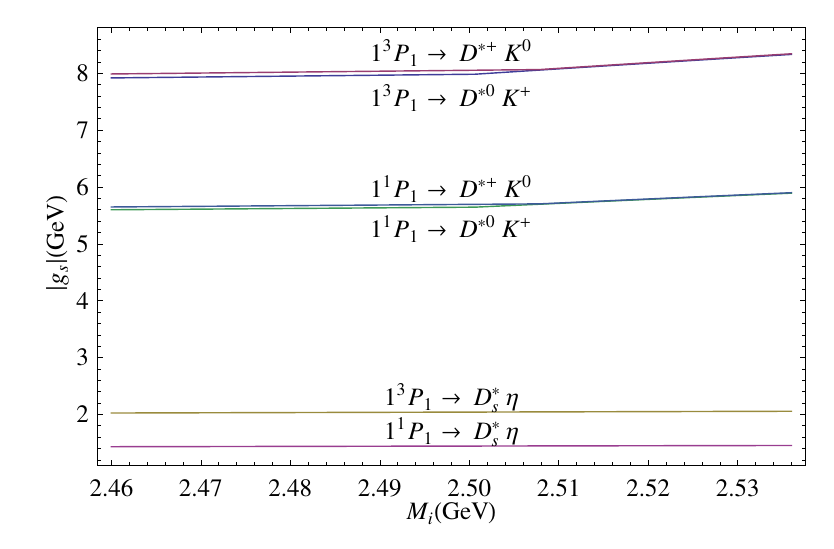}\\
  \vspace{0cm}
  \caption{the absolute values of couplings $g_S$ as functions of the initial meson mass $M_i$}\label{g_s}
\end{figure}

The masses $M_i$ of the initial states $1{}^3 P_1$ and $1{}^1 P_1$
$c \bar{s}$ still have uncertainties. Fortunately, the couplings
$|g_S|$ change only $5\%$ at most when $M_i \in [2.460, \ 2.536] \
\textrm{GeV}$ as shown in Fig.~\ref{g_s}. Also it shows that the
couplings to $D^{*0} K^+$ and $D^{*+} K^0$ are almost the same for
each state due to the isospin symmetry. A set of typical $g_s$
couplings is listed in Table~\ref{tab3}.
\begin{table}[htbp]
  \centering
  \caption{Vertex couplings $g_S$ at $M_i=2.5$ GeV.}\label{tab3}
  \begin{tabular}{c|c|c|c}
    \hline\hline
    $g_S$ (GeV) & $D^{*0} K^+$ & $D^{*+} K^0$ & $D_s^{*} \eta$ \\ \hline
    $1{}^3P_1$ & $-7.982 $   & $-8.052 $   & $2.040 $      \\ \hline
    $1{}^1P_1$ & $5.644  $   & $5.694 $    & $-1.443 $ \\
    \hline\hline
  \end{tabular}
\end{table}

Apart from the on-shell coupling $g_S$, the chiral quark model also
provides an exponential momentum-dependent form factor
$\exp(-{q_1^2}/4 \alpha^2)$ as shown in Eqs.~(\ref{eqn17.1}),
(\ref{eqn17.2}), (\ref{eqn22.1}), and (\ref{eqn22.2}). In order to
keep this feature in the meson loops, we modify the exponential form
factor to a covariant form:
\begin{equation}\label{eqn23}
\exp(-\frac{q_1^2}{4 \alpha^2}) \rightarrow
\exp(\frac{q^2-m^2}{\Lambda^2}) \ ,
\end{equation}
where $q^{\mu}$ and $m$ are the four-vector momentum and mass of
either $\mathbb{V}$ or $\mathbb{P}$ particle. Parameter $\Lambda$ is
the cut-off energy, which can be determined by the quark model,
namely, for the $D^* K$ and $D_s^* \eta$ loops,
\begin{eqnarray}\label{eqn24}
  \Lambda=2 \frac{m_c+m_s}{m_c} \left[\frac{2 m_c}{m_c+m_s}\right]^{\frac{1}{4}} \beta = 1.174 \
  \textrm{GeV} \ .
\end{eqnarray}
The exponential form factor serves to remove the ultraviolet
divergence in the loop integrals.

\section{The propagator matrix}\label{sec:propagator}

In this Section we will determine the propagator matrix $G$. From
Eqs.~(\ref{eqn17.1}) and (\ref{eqn17.2}), we have
\begin{equation}\label{eqn25}
    \Pi_a=-\sqrt{2}\Pi_{ab}, \quad \Pi_b=-\Pi_{ab}/ \sqrt{2}.
\end{equation}
So, we only need to calculate the mixing term $\Pi_{ab}$. With the
AVP coupling form factors, we can explicitly write down $D_{ab}^{\mu
\nu}$ as the following:
\begin{eqnarray}\label{eqn26}
   D_{ab}^{\mu \nu} &=& \sum g_a g_b \cdot i  \int \frac{\ud^4 k}{(2\pi)^4} \frac{ \exp\left(\frac{k^2-m_v^2}{\Lambda^2}\right) \exp\left(\frac{(k+p)^2-m_p^2}{\Lambda^2}\right) }{[k^2-m_v^2][(k+p)^2-m_p^2]}  \left(g^{\mu \nu}-\frac{k^{\mu}k^{\nu}}{m_v^2}\right) \nonumber \\
   &\equiv & \sum g_a g_b(\Pi g^{\mu \nu} + B' p^{\mu}p^{\nu}/p^2) \equiv \sum g_a g_b
   loop \ ,
\end{eqnarray}
where $g_a$ and $g_b$ are the $S$-wave couplings of the two
vertices, respectively. Comparing with Eq.~(\ref{eqn5}), we obtain
\begin{equation}\label{eqn27}
    \Pi_{ab}\equiv \sum g_a g_b \Pi \ .
\end{equation}
The mixing term $\Pi_{ab}$ can be decomposed into two terms, i.e.
\begin{equation}\label{eqn31}
    \Pi_{ab}\equiv\Pi_{ab}^1+\Pi_{ab}^2,
\end{equation}
with
\begin{equation}
\Pi_{ab}^1\equiv \sum g_a g_b \Pi^1, \quad \Pi_{ab}^2\equiv \sum g_a
g_b \Pi^2 \ ,
\end{equation}
where $\Pi_{ab}^1$ and $\Pi_{ab}^2$ are contributions from the
$g^{\mu \nu}$ and $k^{\mu}k^{\nu}$ terms of the vector propagator,
respectively.

As follows, we first make an on-shell approximation to investigate
the absorptive part. Then, we investigate the full integrals with
the help of the exponential form factors.

\subsection{On-shell approximation}

\begin{figure}[htbp]
  \includegraphics[scale=1]{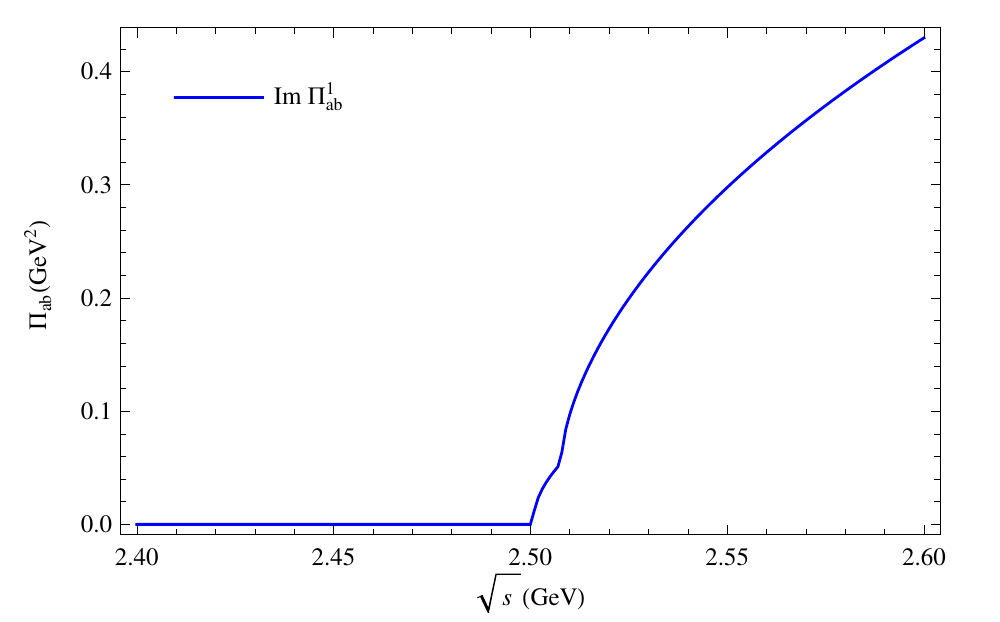}\\
  \vspace{0cm}
  \caption{The mixing term $\Pi_{ab}$ in the on-shell approximation. }\label{Dab}
\end{figure}

Since the absorptive part of a two-point function is independent of
the form factors, the on-shell approximation will allow us to
separate out the absorptive part and then compare it with that in a
full loop integral. Here we only consider $\Pi^1$,
for which the loop integral of Eq.~(\ref{eqn26}) in
the on-shell approximation becomes
\begin{equation}\label{eqn28}
    loop_1 \stackrel{\textrm{on shell}}{\rightarrow} g^{\mu \nu} \frac{-i}{16 \pi^2} \textrm{Im} B0(s,m_p^2,m_v^2) =\Pi^1 g^{\mu
    \nu} \ .
\end{equation}
The resulting mixing term $\Pi_{ab}$ is a function of $s$. We plot
$\Pi_{ab}(\sqrt{s})$ in Fig.~\ref{Dab} with the couplings listed in
Table~\ref{tab3} adopted.

In Fig.~\ref{Dab}, two kink structures can be identified. The first
one at $\sqrt{s}=2.501 \ \textrm{GeV}$ corresponds to the $D^{*0}
K^+$ threshold, and the second one at $\sqrt{s}=2.508 \
\textrm{GeV}$ to the $D^{*+} K^0$ threshold. This result will be
compared with the absorptive part in the full loop integrals later.

\subsection{Full loop calculation with the exponential form factor}

In this Subsection we perform the full loop calculation with the
exponential form factor. The explicit formula for $\Pi^1$ is
\begin{eqnarray}\label{eqn29}
  loop_1&=&
     i\int \frac{\ud^4 k}{(2\pi)^4} \frac{ \exp\left(\frac{k^2-m_v^2}{\Lambda^2}\right) \exp\left(\frac{(k+p)^2-m_p^2}{\Lambda^2}\right) }{[k^2-m_v^2][(k+p)^2-m_p^2]}
     g^{\mu \nu}  \nonumber \\
     &=&
     g^{\mu \nu} \frac{-1}{16 \pi^2} \int_0^1 \ud x \, e^c U(2,1,\frac{b^2}{a},a \Delta)
     =\Pi^1 g^{\mu \nu}
\end{eqnarray}
with
\begin{eqnarray*}
  a      &\equiv & \frac{2}{\Lambda^2} \ ,\\
  b^2    &\equiv & \frac{ (1-2x)^2 }{ \Lambda^4 } s \ , \\
  c      &\equiv & \frac{ s( 2x^2 - 2x + 1 )-m_p^2-m_v^2 }{ \Lambda^2 } \ , \\
  \Delta &\equiv & (1-x)m_v^2 + x m_p^2 -x(1-x)s \ .
\end{eqnarray*}
The explicit formula for $\Pi^2$  is
\begin{eqnarray}\label{eqn32}
    loop_2&=&
    (-i)\int \frac{\ud^4 k}{(2\pi)^4} \frac{ \exp\left(\frac{k^2-m_v^2}{\Lambda^2}\right) \exp\left(\frac{(k+p)^2-m_p^2}{\Lambda^2}\right) }{[k^2-m_v^2][(k+p)^2-m_p^2]}
    \frac{k^{\mu}k^{\nu}}{m_v^2}  \nonumber \\
    &=&
    g^{\mu \nu} \frac{1}{32 \pi^2 a m_v^2} \int_0^1 \ud x \, e^c U(2,0,\frac{b^2}{a},a \Delta) + B_2^U p^{\mu}p^{\nu}  \nonumber \\
    &\equiv &
    \Pi^2  g^{\mu \nu} + B_2 p^{\mu}p^{\nu} \ ,
\end{eqnarray}
where $a, \ b, \ c, \ \Delta$ are the same as those in
Eq.~(\ref{eqn29}). The function $U(a,b,c,z)$ is a class of special
integrals which appears in the evaluation of the loop integrals with
exponential form factors. The detailed definition and calculation of
$U(a,b,c,z)$ are provided in Appendix~\ref{app:1}.

\begin{figure}[htbp]
  \includegraphics[scale=1]{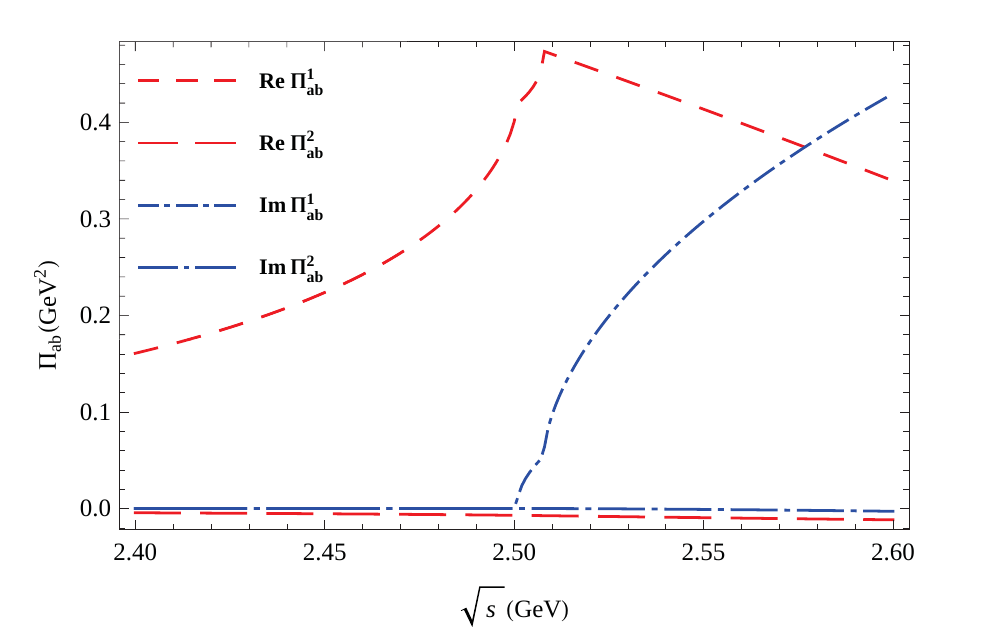}\\
  \vspace{0cm}
  \caption{(color online). The mixing term $\Pi_{ab}$ with exponential form factors.
  $\Pi_{ab}^1$ and $\Pi_{ab}^2$ are the contributions from the $g^{\mu \nu}$ term and $k^{\mu}k^{\nu}$ term of the vector propagator, respectively.
  The dashed lines represent the dispersive  parts, while the dot-dashed lines represent the absorptive ones.}\label{DabExp}
\end{figure}

The full loop calculation of the mixing term $\Pi_{ab}(\sqrt{s})$ is
presented in Fig.~\ref{DabExp} where the parameters are the same as
before. In order to see clearly the contributions from different
parts, we also give two sets of the calculated values in
Tables~\ref{tab:2460} and \ref{tab:2536}.
\begin{table}[htbp]
  \centering
  \caption{The mixing term $\Pi_{ab}$ at the pole position $\sqrt{s}=2.4545 \ \textrm{GeV}$. }\label{tab:2460}
  \begin{tabular}{c||c|c|c||c}
    \hline \hline
    intermediate state & $D^{*0}K^+$ & $D^{*+} K^0$ & $D_s^* \eta$ & $\Pi_{ab}=\sum g_a g_b \Pi \ ({\textrm{GeV}}^2)$ \\ \hline
    $g_a g_b ({\textrm{GeV}}^2)$       & $-45.05$ & $-53.90$ & $-2.944$& ---\\ \hline
    $\Pi^1$(on-shell) & $0$ & $0$ & $0$& $0$\\ \hline
    $\Pi^1$         & $-2.391\times 10^{-3}$ & $-2.253 \times 10^{-3}$ & $-0.868 \times 10^{-3}$ & $0.2317$ \\ \hline
    $\Pi^2$      & $5.650 \times 10^{-5}$ & $5.448 \times 10^{-5}$ & $2.503 \times 10^{-5}$ & $-0.0056$ \\
    \hline \hline
  \end{tabular}
\end{table}
\begin{table}[htbp]
  \centering
  \caption{The mixing term $\Pi_{ab}$ at the pole position $\sqrt{s}=(2.5449-0.0010 i)\ \textrm{GeV}$.}\label{tab:2536}
  \begin{tabular}{c||c|c|c||c}
    \hline \hline
    intermediate state & $D^{*0}K^+$ & $D^{*+} K^0$ & $D_s^* \eta$ & $\Pi_{ab}=\sum g_a g_b \Pi  \ ({\textrm{GeV}}^2)$ \\ \hline
    $g_a g_b ({\textrm{GeV}}^2)$       & $-45.05$ & $-53.90$ & $-2.944$& ---\\ \hline
    $\Pi^1$(on-shell) & $-2.969i\times 10^{-3}$ & $-2.718 i\times 10^{-3}$ & $0$ & $0.2803i$ \\ \hline
    $\Pi^1$         & $(-4.157-2.969i)\times 10^{-3}$ & $(-4.260-2.718i)\times 10^{-3}$ & $-1.389\times 10^{-3}$ & $0.4210+0.2803i$ \\ \hline
    $\Pi^2$      & $(9.419+0.886i) \times 10^{-5}$ & $(9.059+0.678i )\times 10^{-5}$ & $3.560 \times 10^{-5}$ & $-0.0092-0.0008i$ \\ \hline
    \hline
  \end{tabular}
\end{table}

The loop calculation results help us to learn the following points:

\begin{itemize}
\item The imaginary part of $\Pi_{ab}^1$ with exponential form factors
is the same as that in the on-shell approximation. It justifies our
calculation method for $U(a,b,c,z)$ as described in the Appendix.

\item The contribution from the term of $g^{\mu \nu}$ is dominant. The open thresholds
of $D^{*0} K^+$ and $D^{*+}K^0$ cause two kinks in both real and
imaginary parts. With the increase of $\sqrt{s}$, $\textrm{Re}
\Pi_{ab}^1$ first increases until it reaches a summit at the $D^{*+}
K^0$ threshold. It then decreases in a linear behavior in terms of
$\sqrt{s}$. In contrast, $\textrm{Im} \Pi_{ab}^1$ is zero below the
$D^{*0} K^+$ threshold and then increases quickly when the decay
thresholds are open. One can see that below the $D^{*0} K^+$
threshold, the real part is the only contribution and cannot be
neglected. The imaginary part becomes significant above 2.53 GeV.

\item The calculation also shows that the contributions from the $k^{\mu} k^{\nu}$
term of the vector propagator are negligible. Near the threshold,
the momentum is small such that $\Pi^2_{ab}$ suffers an $O(1/m_v^2)$
suppression comparing to $\Pi^1_{ab}$ in both the absorptive and
dispersive part.

\item The contributions from the $D^* K$ loops are found dominant, while the
contributions from $D_s^* \eta$ account for only about $1\%$ of the
mixing term due to the rather small coupling value in the $D_s^*
\eta$ loop.

\end{itemize}

\section{Pole positions and mixing parameters }\label{sec:mass}

\begin{figure}[htbp]
  \includegraphics[scale=1]{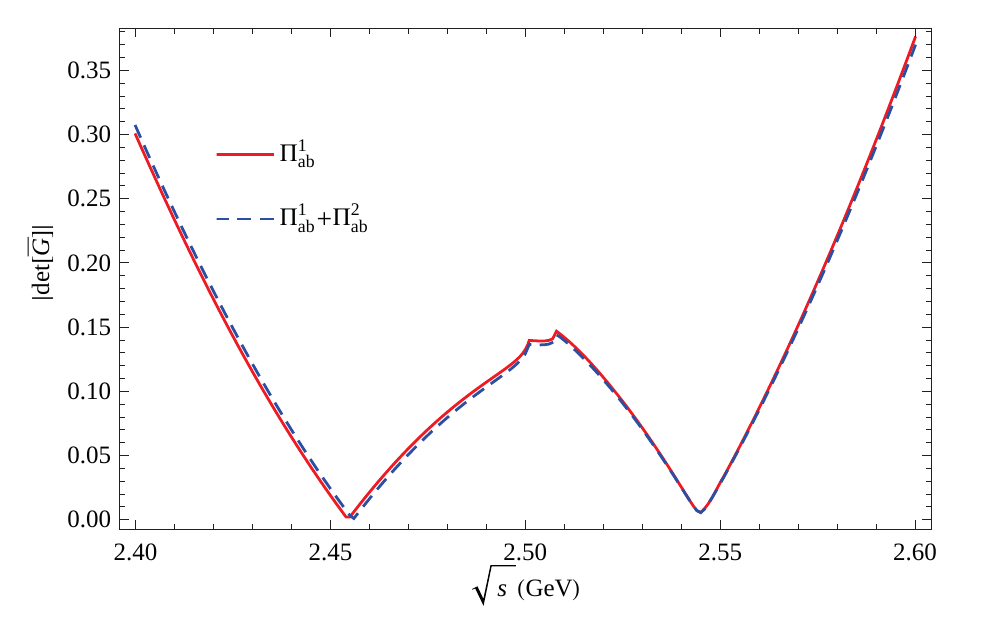}\\
  \vspace{0cm}
  \caption{Pole structures highlighted by the zero values of $\det[\bar{G}]$ in the propagator matrix. }\label{poles}.
\end{figure}

With the $\Pi_{ab}(s)$ determined, we can directly search for poles
for the physical states in the propagator matrix $G$ in
Eq.~(\ref{eqn6}). We adopt the following bare $c\bar{s}$ masses,
$m[{}^3P_1]=2.57 \ \textrm{GeV}$ and $m[{}^1P_1]=2.53 \
\textrm{GeV}$, from the Godfrey-Isgur (GI)
model~\cite{Godfrey:1985xj} as input. By scanning over the energy
$\sqrt{s}$, the requirement of $|\det[\bar{G}(s)]|=0$ provides a
direct access to the pole positions as shown in Fig.~\ref{poles}.
Two possible poles near 2.46 GeV and 2.54 GeV are highlighted. When
varying the cut-off parameter $\Lambda$ in Eq.~(\ref{eqn24}) within
the range of $[1.174-0.22,1.174+0.22]$ GeV , it shows that the
higher pole is stable and the lower one changes from 2.47 GeV to
2.44 GeV. Searching for the poles on the complex energy plane, we
can pin down the masses and widths of these two poles as listed in
Table~\ref{tab:mass}. It shows that the mass of $D_{s1}(2460)$
changes 3.6 MeV at most with or without the contribution from the
$k^{\mu}k^{\nu}$ term of the propagator, while the mass of
$D_{s1}(2536)$ changes only 0.1 MeV. The extracted mass of
$D_{s1}(2460)$ is only 5 MeV below the experiment value, and the
mass of $D_{s1}(2536)$ is only 10 MeV above the experiment one. In
principle, the Okubo-Zweig-Iizuka (OZI) rule allowed hadronic decay
width of $D_{s1}(2460)$ is zero. The obtained width 2.0 MeV for
$D_{s1}(2536)$ seems to be slightly larger than the experiment value
0.92 MeV, but can still be regarded as in good agreement. In brief,
our prediction for the masses and widths of these two states agrees
well with the experiment data.

\begin{table}[htbp]
  \centering
  \caption{Masses and widths obtained from the pole analysis.}\label{tab:mass}
  \begin{tabular}{c|c|c}
    \hline \hline
    $[m-i\frac{\Gamma}{2}]$ (MeV) & $D_{s1}(2460)$ & $D_{s1}(2536)$ \\ \hline
    $\Pi_{ab}^1              $ & $2454.5$       & $2544.9-1.0 i$ \\ \hline
    $\Pi_{ab}^1+\Pi_{ab}^2$ & $2455.8$       & $2544.9-1.1 i$ \\ \hline
    Experiment                 & $2459.5$       & $2535.08-0.46 i$ \\
    \hline \hline
  \end{tabular}
\end{table}

Before extracting the mixing parameters, we show that our formalisms
can reproduce the ideal mixing angle $\theta_0$ in the heavy quark
limit. In this limit, $m_a$ and $m_b$ are degenerate. From
Eqs.~(\ref{eqn7}) and (\ref{eqn25}), we only need to diagonalize the
simple matrix
\begin{equation}\label{eqn33}
\left(
  \begin{array}{cc}
    -\frac{1}{\sqrt{2}} & -1 \\
    -1 & -\sqrt{2} \\
  \end{array}
\right) \ ,
\end{equation}
which immediately leads to
$\theta_0=\arctan[1/\sqrt{2}]=35.26^\circ$.

Now we proceed to the extraction of the mixing parameters
$\{\theta,\phi\}$ by diagonalizing $\bar{G}(s)$ with $\sqrt{s}$
fixed at the poles. When $\bar{G}$ is a complex matrix, we try to
approach the diagonal limit $R \bar{G}_{ab}R^{\dagger}=\bar{G}_{AB}$
in three ways: Method \Rmnum{1}, set $\bar{G}_{12}^{AB}=0$; Method
\Rmnum{2}, set $\bar{G}_{21}^{AB}=0$; and Method \Rmnum{3}, minimize
$|\bar{G}_{12}^{AB}| + |\bar{G}_{21}^{AB}|$. The results from these
three diagonalization schemes are listed in Table~\ref{tab:mixing}.
As we expected before, the mixing angles of these two states
determined at their pole masses are indeed different. For
$D_{s1}(2460)$, $\bar{G}$ is a symmetric real matrix. So the mixing
parameters are the same in these three methods: $\theta=47.6^\circ,
\, \phi=0^\circ$. From the mixing scheme in Eq.~(\ref{eqn2:1}),
$\theta>45^\circ$ means that the ${}^1P_1$ component is larger than
the ${}^3P_1$ in $D_{s1}(2460)$. This mixing pattern would affect
the mass shift as we will show later. The result corresponds to
$\theta'=12.3^\circ$ in the $j=1/2$ and $j=3/2$ mixing in the heavy
quark limit. For $D_{s1}(2536)$ , $\bar{G}$ is a complex matrix. The
mixing angle $\theta=39.7^\circ$ determined at the $D_{s1}(2536)$
mass changes little in those three methods, while the relative phase
suffers an uncertainty of $\phi=-6.9^\circ \sim 6.9^\circ$. We will
show later in Sec.~\ref{sec:exp} that the mixing angle
$\theta=39.7^\circ$ is consistent with the experimental constraints
and can be useful for picking up one of those two solutions from the
experimental fit. Again from the mixing scheme, $\theta<45^\circ$
means that the ${}^1P_1$ component is larger than the ${}^3P_1$ one
in $D_{s1}(2536)$. The result corresponds to $\theta'=4.4^\circ$ in
the $j=1/2$ and $j=3/2$ mixing bases. The energy dependence of the
mixing angle reflects the breaking of orthogonality among these two
physical states.

\begin{table}[htbp]
  \centering
  \caption{The mixing angle $\theta$ and relative phase $\phi$ extracted at the two poles in those three diagonalization schemes.}\label{tab:mixing}
  \begin{tabular}{|c||c|c|c||c|c|c|}
    \hline
      & \multicolumn{3}{c||}{$D_{s1}(2460)$} & \multicolumn{3}{c|}{$D_{s1}(2536)$} \\ \cline{2-7}
    \raisebox{2.3ex}[0pt]{$\{\theta,\phi\}[{}^{\circ}]$}& \Rmnum{1} & \Rmnum{2} & \Rmnum{3} & \Rmnum{1} & \Rmnum{2} & \Rmnum{3} \\ \hline
    $\Pi_{ab}^1              $ & $\{47.5, \ 0\}$ & $\{47.5, \ 0\}$ & $\{47.5, \ 0\}$ & $\{39.7, \ -6.4\}$ & $\{39.7, \ 6.4\}$ & $\{39.7, \ 0\}$ \\ \hline
    $\Pi_{ab}^1+\Pi_{ab}^2$ & $\{47.6, \ 0\}$ & $\{47.6, \ 0\}$ & $\{47.6, \ 0\}$ & $\{39.8, \ -6.5\}$ & $\{39.8, \ 6.5\}$ & $\{39.7, \  0\}$ \\
    \hline
  \end{tabular}
\end{table}

From the mixing angle analysis, we also learn that the
$D_{s1}(2460)$ has a larger $j=1/2$ component which couples to the
$D^* K$ through an $S$-wave. It hence acquires a significant mass
shift $\sim 100$ MeV through meson loop corrections. In contrast,
the $D_{s1}(2536)$ contains a larger $j=3/2$ component which couples
to the $D^*K$ through a $D$-wave. It only gains a small mass shift
$\sim 10$ MeV.
\begin{figure}[hbtp]
  \includegraphics[scale=1]{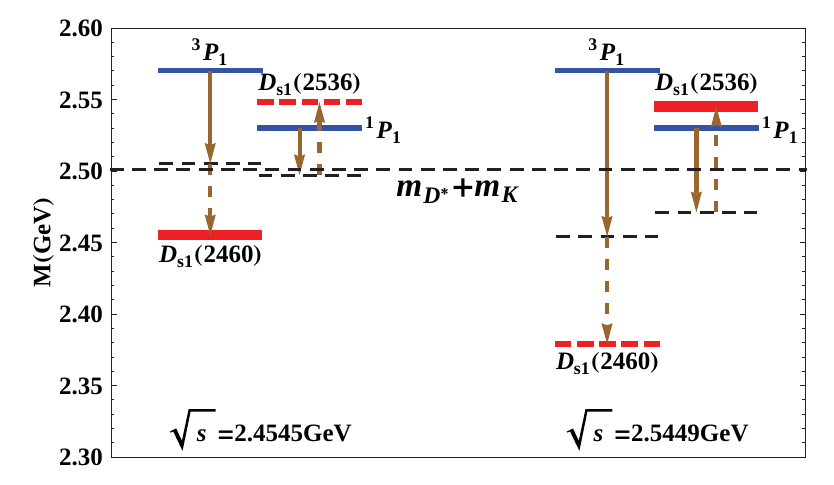}
  \vspace{0cm}
\caption{(color online). Schematic plot for the mass-shift
procedure. The thin solid bars represent the original ${}^3P_1$ and
${}^1P_1$ states in the quark model. The thick solid bars represent
the two physical states $D_{s1}(2460)$ (left) and $D_{s1}(2536)$
(right). The solid arrows represent the mass shifts due to the
diagonal elements $\Pi_a$ and $\Pi_b$, while the dashed arrows
represent those due to the off-diagonal element $\Pi_{ab}$. The
threshold for $D^* K$ is shown by the horizontal dashed
line.}\label{massShift}
\end{figure}
\begin{table}[htbp]
  \centering
  \caption{Mass shift procedure at different $\sqrt{s}$. From $\sqrt{M_{ab}^2}$,  we can see the mass shifts
  due to $\Pi_a$ and $\Pi_b$, while from $\sqrt{M_{AB}^2}$, further  mass shifts due to $\Pi_{ab}$ can be learned.
  Here we only present the results from the quasi-diagonalization method \Rmnum{2}. }\label{tab:shift}
  \begin{tabular}{cccc}
    \hline \hline
    $\sqrt{s}$ GeV          & bare mass (GeV)    &  $\sqrt{M_{ab}^2}$ (GeV) & $\sqrt{M_{AB}^2}$ (GeV) \\ \hline
    2.4545 GeV &
    & $\left[\begin{array}{cc}
                      2.497   & 0.481 i \\
                      0.481 i & 2.505
    \end{array}\right]$
    & $\left[\begin{array}{cc}
                      2.548   & 0 \\
                      0       & 2.455
                    \end{array}
    \right]$
\\
    2.5449 GeV &
    \raisebox{3ex}[-10pt]
    {$\left[\begin{array}{cc}
                             2.53 & 0 \\
                             0    & 2.57
                           \end{array}\right]$}
    & $\left[\begin{array}{cc}
            2.471-0.040i  & 0.206-0.678i \\
            0.206-0.678i  & 2.454-0.080i
          \end{array}
    \right]$
    & $\left[\begin{array}{cc}
              2.545-0.001i & 0.141+0.303i \\
              0            & 2.379-0.123i
            \end{array}
    \right]$ \\
    \hline \hline
  \end{tabular}
\end{table}

The mass shift procedure is also an interesting issue and can help
us to understand why $D_{s1}(2460)$ has a larger ${}^1P_1$
component. As shown in Fig.~\ref{massShift} and
Table~\ref{tab:shift}, we can decompose the mass shift procedure
into two classes, i.e. diagonal shift and off-diagonal shift. The
diagonal elements $\Pi_a$ and $\Pi_b$ cause both ${}^3P_1$ and
${}^1P_1$ states to move downwards, while the off-diagonal elements
$\Pi_{ab}$ make one state to shift up and the other to shift down.
At $\sqrt{s}=2.46 \ \textrm{GeV}$, after the diagonal shift the
${}^3P_1$ state is still higher than the ${}^1P_1$. But after the
off-diagonal shift, the higher mass state moves down to become an
on-shell $D_{s1}(2460)$ and the lower state moves up to become a
virtual $D_{s1}(2536)$. The reversal of the mass ordering results in
a mixing angle $\theta > 45^\circ$ and thus a larger ${}^1P_1$
component in $D_{s1}(2460)$. At $\sqrt{s}=2.54 \ \textrm{GeV}$,
after the diagonal shift the ${}^1P_1$ becomes higher than the
${}^3P_1$. Then after the off-diagonal shift, the higher state
becomes much higher and the lower much lower, which causes a mixing
angle $\theta < 45^\circ$ and a larger ${}^1P_1$ component in
$D_{s1}(2536)$. Note that in this situation the on-shell state
corresponds to the $D_{s1}(2536)$, and the $D_{s1}(2460)$ appear as
a virtual one.

\section{Experimental constraints on the mixing angle}\label{sec:exp}

In this part,  we come to survey the constraints for the mixing
angle $\theta$ from experiments. The strong decays of $D_{s1}(2536)$
has been measured with reasonable precision which are summarized in
Table~\ref{tab:expdata}. Since the $D^* K$ channel is the only
allowed strong decay channel for $D_{s1}(2536)$, it is a good
approximation to assume
\begin{equation}\label{eqn:exp1}
    \Gamma[D_{s1}(2536)] \approx \Gamma( D_{s1}(2536)\to D^* K ) \ ,
\end{equation}
which can be estimated in the chiral quark model. The partial width
fractions $R_1$ and $R_2$ can also be calculated and compared with
the data.

\begin{table}[htbp]
  \centering
  \caption{ The available experimental status of $D_{s1}(2536)$. }\label{tab:expdata}
  \begin{tabular}{c}
    \hline\hline
    $
      m = 2535.08\pm 0.01 \pm 0.15 \ \textrm{MeV}, \quad
      \Gamma = 0.92 \pm 0.03 \pm 0.04 \ \textrm{MeV} \quad (\textrm{BaBar \cite{Lees:2011um}})
    $
    \\ \hline
    $
      R_1 = \frac{ \Gamma (D^*(2007)^0 K^+ ) }{ \Gamma (D^*(2010)^+ K^0 ) } = 1.36 \pm 0.20  \quad (\textrm{PDG2010 \cite{Nakamura:2010zzi}})
    $
    \\ \hline
    $
      R_2 = \frac{ \Gamma (D^*(2010)^+ K^0 )_{S-wave} }{ \Gamma (D^*(2010)^+ K^0 ) } = 0.72 \pm 0.05 \pm 0.01  \quad(\textrm{Belle \cite{:2007dya}})
    $
    \\
    \hline\hline
  \end{tabular}
\end{table}

\begin{figure}[htbp]
  \begin{minipage}[b]{0.33\textwidth}
    \includegraphics[width=0.9\textwidth]{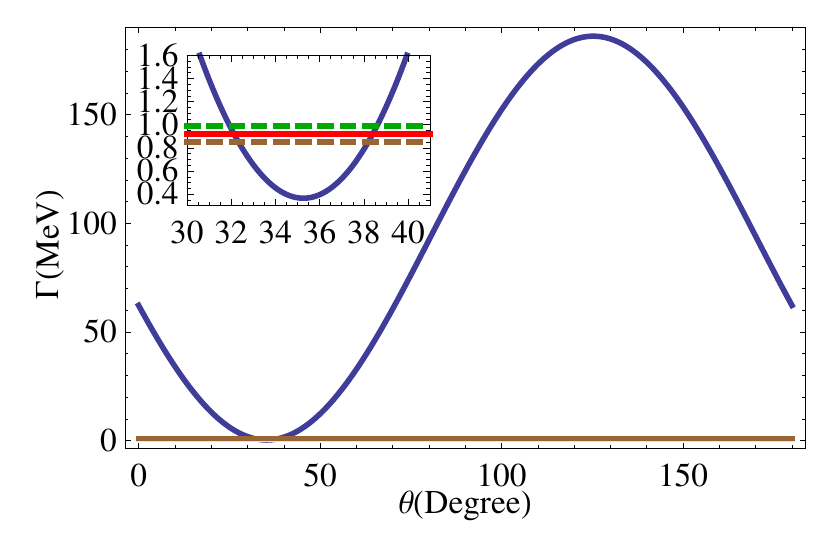}\\
    \caption{$\Gamma[D_{s1}(2536)]$ as a function of $\theta$}\label{fig:width}
  \end{minipage}%
  \begin{minipage}[b]{0.33\textwidth}
    \includegraphics[width=0.9\textwidth]{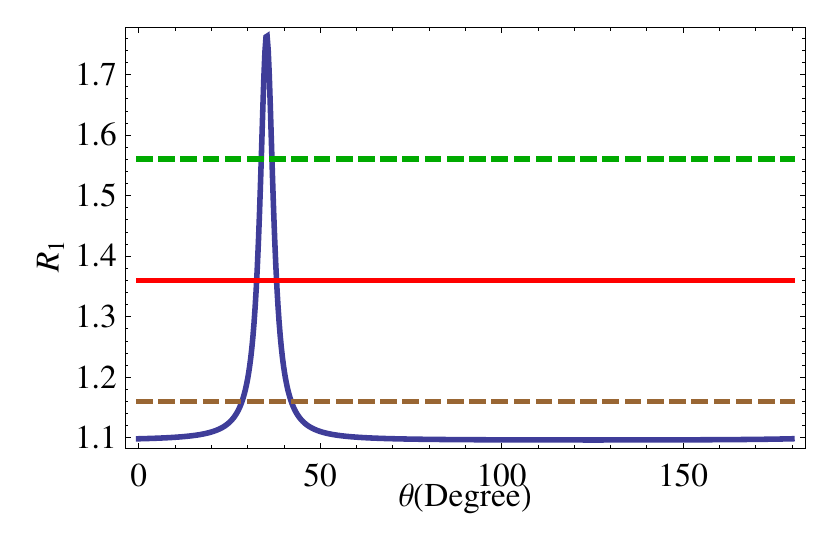}\\
    \caption{$R_1$ as a function of $\theta$}\label{fig:R_1}
  \end{minipage}%
  \begin{minipage}[b]{0.33\textwidth}
    \includegraphics[width=0.9\textwidth]{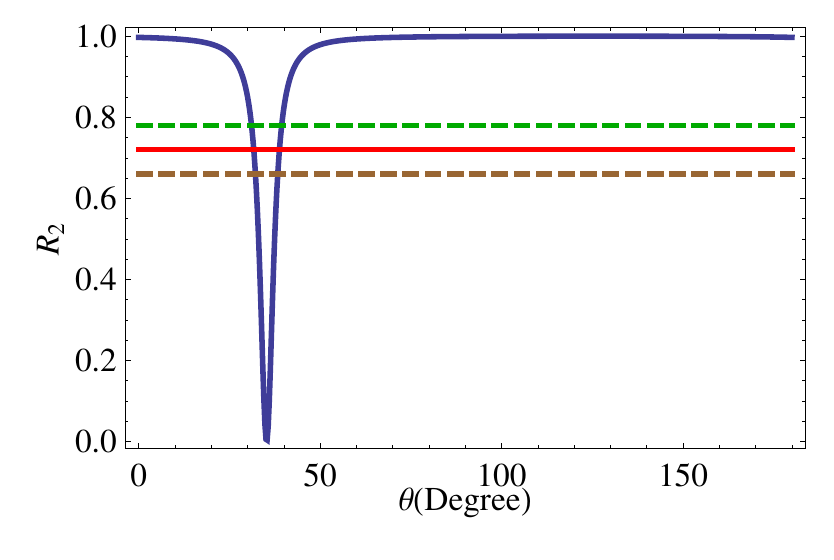}\\
    \caption{$R_2$ as a function of $\theta$}\label{fig:R_2}
  \end{minipage}
\end{figure}

\begin{figure}[hbtp]
  \includegraphics[scale=1]{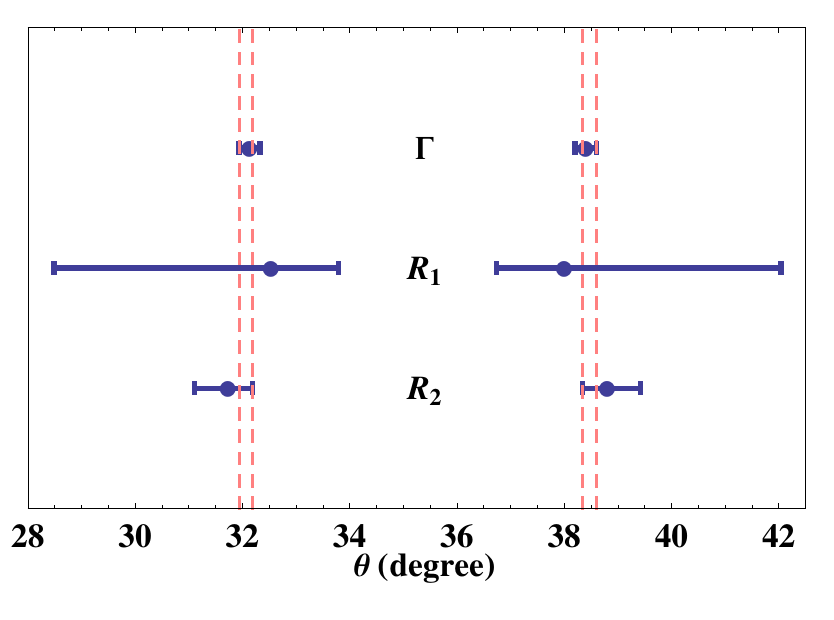}
  \vspace{0cm}
\caption{The experimental constraints on the mixing angle $\theta$.
}\label{fig:exp_theta}
\end{figure}

The helicity amplitudes for $ 1{}^3P_1 \to D^* K$ and $1{}^1P_1 \to
D^* K$ have been listed in Eqs.~(\ref{eqn14}) and (\ref{eqn15}). The
partial width can be obtained by~\cite{Zhong:2008kd}
\begin{equation}\label{eqn:exp2}
    \Gamma =\left( \frac{\delta}{f_m} \right)^2 \frac{( E_f + M_f ) |\vec{q}| }{ 4 \pi M_i ( 2J_i + 1 ) } \sum_{\nu} \left| A_{\nu}^q
    \right|^2 \ ,
\end{equation}
where $J_i$ is the spin of the initial particle. In order to
calculate $R_2$, we need to extract the $S$-wave components from the
helicity amplitudes. By defining $A_s= \frac{G_S}{ 2\sqrt{\pi} }$
and $A_D= \frac{G_D}{ 2\sqrt{\pi} }$, we deduce from
Eq.~(\ref{eqn10})
\begin{equation}\label{eqn:exp3}
    \left\{\begin{array}{l}
             A_0 = A_S - \sqrt{2} A_D \\
             A_1 = A_S + \frac{1}{\sqrt{2}} A_D
           \end{array}
    \right.
    \Rightarrow
    \left\{\begin{array}{l}
             A_S = \frac{1}{3}( A_0 + 2 A_1 ) \\
             A_D = -\frac{\sqrt{2}}{3}( A_0 - A_1 )
           \end{array}
    \right. \ ,
\end{equation}
where the $S$ and $D$-wave components have been separated out. We
use the same model parameters as before to calculate the partial
width $\Gamma[D_{s1}(2536)]$ and ratios $R_1$ and $R_2$ in terms of
the mixing angle $\theta$. The results are shown in
Figs.~(\ref{fig:width})-(\ref{fig:exp_theta}).

A similar result as Fig.~\ref{fig:width} for  $\Gamma[D_{s1}(2536)]$
in terms of $\theta$ has been given in Ref.~\cite{Zhong:2008kd} but
with the notation $\theta \to \phi + 90^\circ$. Those three
horizontal lines in Figs.~(\ref{fig:width})-(\ref{fig:R_2})
represent the upper limits, center values, and lower limits of the
experimental data. The interesting feature arising from the results
of Figs.~(\ref{fig:width})-(\ref{fig:R_2}) is that the overlaps
between the experimental data and theoretical values are separated
into two narrow bands of $\theta$ which are located symmetric to the
ideal mixing angle $\theta_0=35.26^\circ$, i.e. $\theta_1\simeq
32.1^\circ$ or $\theta_2\simeq 38.4^\circ$. An alternative way to
present the results is via Fig.~\ref{fig:exp_theta}, where the
overlapped $\theta$ values are denoted by the vertical dashed lines,
while the experimental observables with errors are presented in
terms of $\theta$. Notice that these two bands of $\theta$ are both
smaller than $45^\circ$. Therefore, based on the present
experimental measurements, one cannot determine which value for
$\theta$ should be taken. It turns out that our analysis in
Sec.~\ref{sec:mass} can precisely pick up one of these two
solutions, namely, $\theta_2\simeq 38.4^\circ$ is favored in
comparison with the theoretical value $\theta=39.7^\circ$.

\section{summary}\label{sec:summary}

In summary,  we have studied the mixing mechanism for the axial
vector states $D_{s1}(2460)$ and $D_{s1}(2536)$ via the $S$-wave
intermediate meson loops. We establish the propagator matrix for
this two-state system. Then, by searching for the pole structures in
the propagator matrix, we can pin down the masses and widths of the
physical states. The mixing angle and relative phase between the
${}^3P_1$ and ${}^1P_1$ components can be determined by
diagonalizing the propagator matrix. For $D_{s1}(2460)$, we obtain
$m=2454.5 \ \textrm{MeV},\, \theta=47.5^\circ$ and $\phi=0^\circ$.
For $D_{s1}(2536)$, we find $m=2544.9-1.0 i \ \textrm{MeV},\,
\theta=39.7^\circ$, and $\phi=-6.9^\circ \sim 6.9^\circ$. Our
results agree well with the experimental measurement. In particular,
the new BaBar measurement put a strong constraint on the mixing
angle at the mass of $D_{s1}(2536)$ with two solutions,
$\theta_1\simeq 32.1^\circ$ and $\theta_2\simeq 38.4^\circ$. Our
theoretical calculation finds $\theta=39.7^\circ$ which is in good
agreement with $\theta_2$.

Note that due to the breaking of orthogonality the energy-dependent
mixing angles defined at the different physical masses turn out to
have different values. We find that both $D_{s1}(2460)$ and
$D_{s1}(2536)$ have a relatively large ${}^1P_1$ component in their
wavefunctions.

It is also interesting to learn the important role played by the
coupled channel effects for states near open thresholds. For states
that can couple to each other via the coupled channels, the
two-state propagator matrix carries rich information about the
mixing and mass shifts as a manifestation of the underlying
dynamics. Extension of such a study to other axial-vector meson
mixings would be useful for deepen our understanding of the coupled
channel effects and their impact on the hadron spectrum.

\acknowledgments

This work is supported, in part, by National Natural Science
Foundation of China (Grant No. 11035006), Chinese Academy of
Sciences (KJCX2-EW-N01), and Ministry of Science and Technology of
China (2009CB825200).

\appendix

\section{Calculation of function $U(a,b,c,z)$}\label{app:1}

\begin{figure}[htbp]
  \includegraphics[scale=0.6]{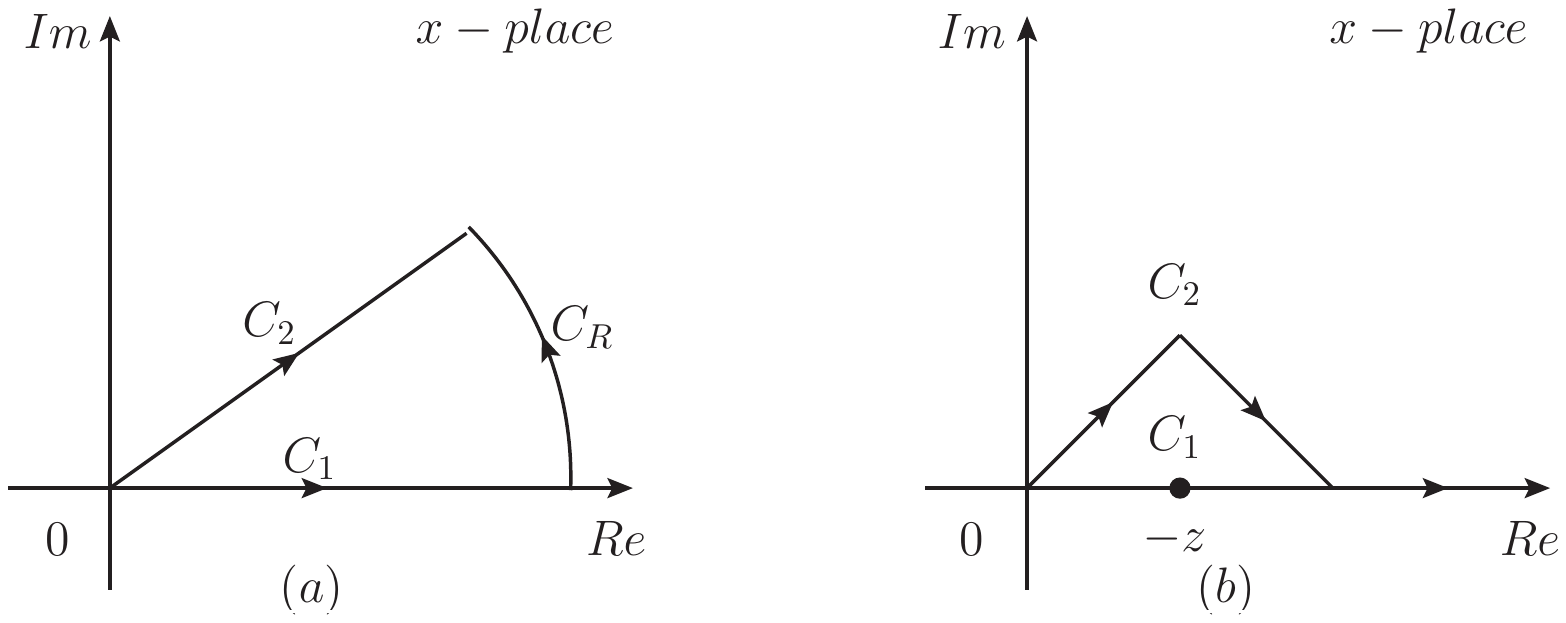}\\
  \vspace{-12cm}
  \caption{The continuation of U(a,b,c,z)}\label{continuation}
\end{figure}

Initially we define
\begin{equation}\label{eqn:app:1}
    U(a,b,c,z)=\frac{1}{\Gamma(a)} \int_0^\infty \ud t \ t^{a-1} (1+t)^{b-a-1} \exp\left(-z t
    -\frac{c}{1+t}\right) \ ,
\end{equation}
which is the typical integral we encounter in the calculation. A
special case, $U(a,b,c=0,z)=U(a,b,z)$, is the Tricomi confluent
hypergeometric function, which is a build-in function in
Mathematica. The function $U(a,b,z)$ is a single-valued function on
the $z$-plane cut along the interval $(-\infty, 0]$, where it is
continuous from above, i.e.
\begin{eqnarray}\label{eqn:app:2}
    \textrm{when}\quad z<0, & & U(a,b,z)=\lim_{\epsilon \to 0^+} U(a,b,z+i
    \epsilon) \ .
\end{eqnarray}
Function $U(a,b,c,z)$ as a physical quantity should be analytic with
respect to its arguments. However, the integral in
Eq.~(\ref{eqn:app:1}) only converges when $Re(z)>0$ and $Re(a)>0$.
In order to analytically continue the integral to $Re(z)<0$, we make
a change in variables $z t= x (z>0)$. Hence, Eq.~(\ref{eqn:app:1})
becomes
\begin{equation}\label{eqn:app:3}
U(a,b,c,z)= \frac{z^{1-b}}{\Gamma(a)} \int_0^{\infty} \ud x \
x^{a-1} (x+z)^{b-a-1} \exp\left(-x -\frac{c z}{ x + z }\right) \ .
\end{equation}
In the region $Re(z)>0$, Eq.~(\ref{eqn:app:1}) and (\ref{eqn:app:3})
are exactly equivalent to each other when $Re(b)<2$. The difference
between them can be expressed by the integral over $C_R$ in
Fig.~\ref{continuation}(a).

\noindent
When $Re(b)<2$, the contribution from $C_R$ is zero. Comparing with
Eq.~(\ref{eqn:app:1}), the integral in Eq.~(\ref{eqn:app:3}) has
larger convergent region, i.e. the whole complex $z$-plane except
$z<0$. When $z<0$, there is a singular point at $x=-z$ in the
integral path as shown in Fig.~\ref{continuation}(b). Considering
Eq.~(\ref{eqn:app:2}), $U(a,b,c,z)$ must satisfy a similar
requirement. It means that the integral path $C_1$ in
Fig.~\ref{continuation}(b) should be replaced by the integral path
$C_2$. Using the expression in Eq.~(\ref{eqn:app:3}) and the
replacement in Fig.~\ref{continuation}(b), we can analytically
continue the integral in Eq.~(\ref{eqn:app:1}) to the whole
$z$-plane. The constraints of the above method are $Re(a)>0$ and
$Re(b)<2$, which could satisfy our present need.

\begin{figure}[htbp]
\begin{minipage}[l]{0.45\textwidth}
  \centering
  \includegraphics[scale=0.8]{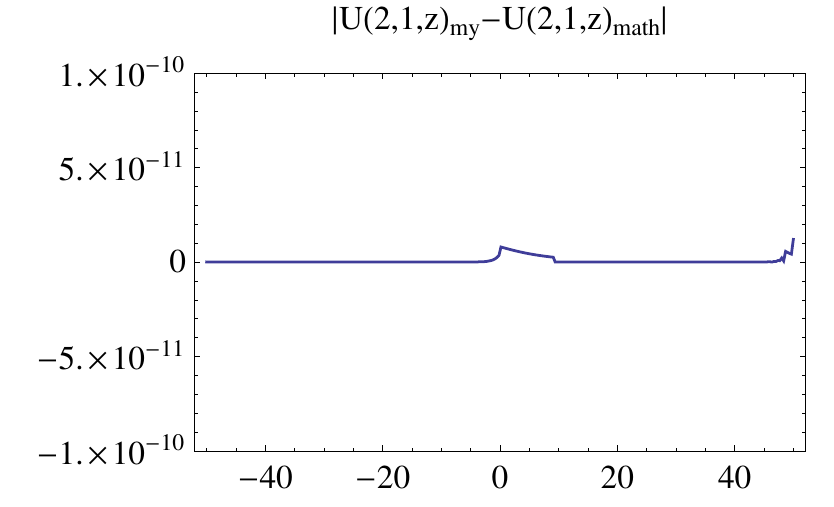}
  \vspace{0cm}
\end{minipage}
\begin{minipage}[l]{0.45\textwidth}
  \centering
  \includegraphics[scale=0.8]{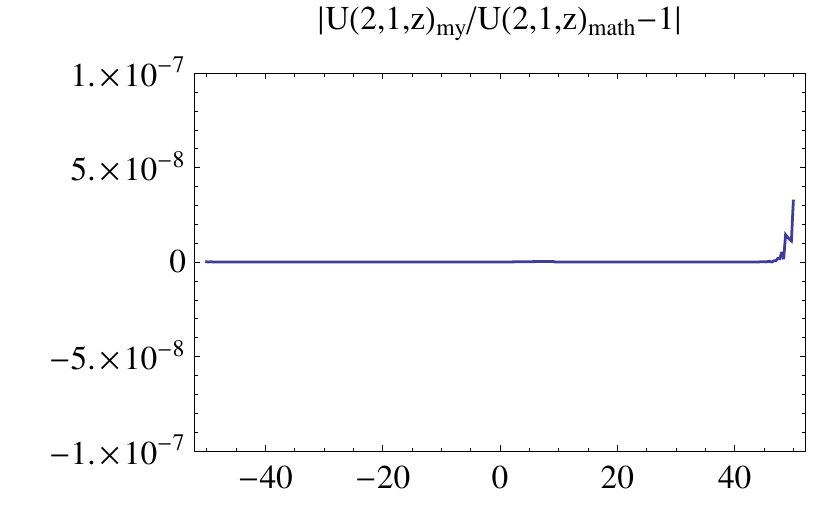}
  \vspace{0cm}
\end{minipage}
\begin{minipage}[l]{0.45\textwidth}
  \centering
  \includegraphics[scale=0.8]{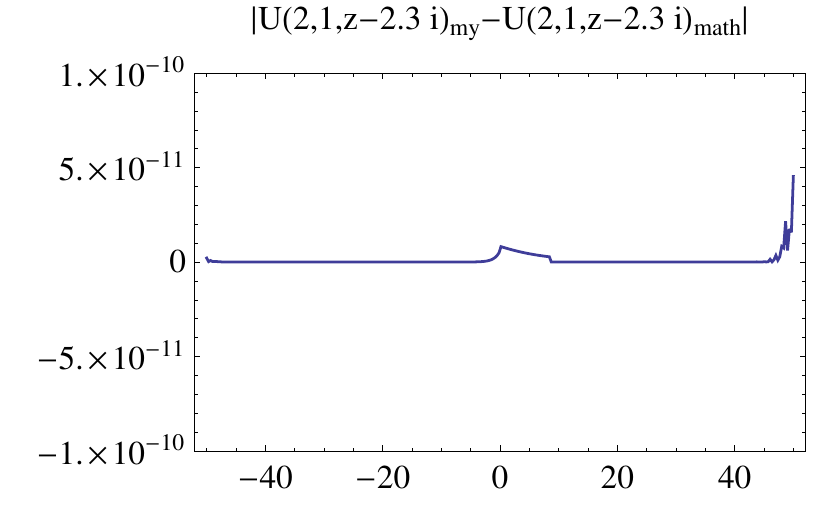}
  \vspace{0cm}
\end{minipage}
\begin{minipage}[l]{0.45\textwidth}
  \centering
  \includegraphics[scale=0.8]{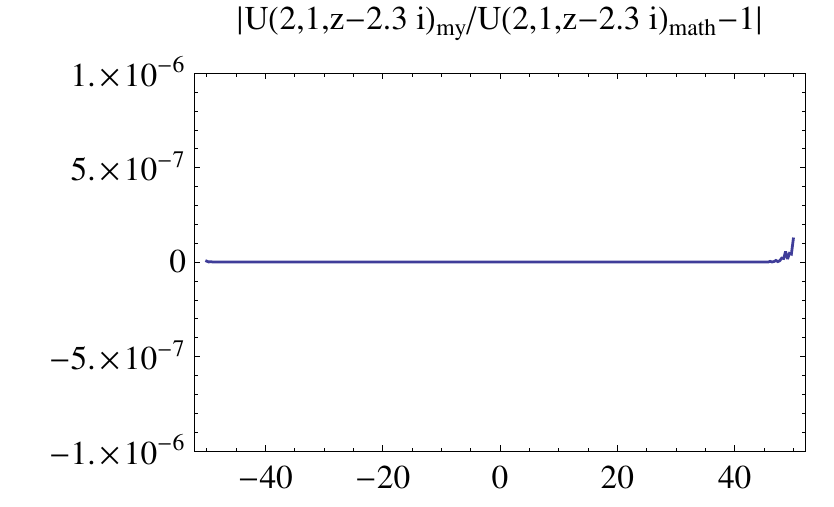}
  \vspace{0cm}
\end{minipage}
  \caption{Test the accuracy and precision of the analytic continuation of $U(a,b,c,z)$.}\label{pic_test}
\end{figure}

To test this method, we compare the results for $U(a,b,z)$ in
Fig.~\ref{pic_test} using our method and the build-in Mathematica
program. It shows that these two calculations are in good agreement
to each other. This test is done at $c=0$. Since $c$ in
Eq.~(\ref{eqn:app:3}) does not bring either new divergence problems
or new singular points, we can justify that our analytic
continuation of $U(a,b,c,z)$ is quite reliable and generally
applicable.

\bibliography{reference}

\end{document}